\documentclass[article,shortnames]{jss}

\usepackage{Sweave}
\usepackage{graphicx}
\usepackage{caption}
\author{Xiyue Liao\\Colorado State University\And Mary C. Meyer\\Colorado State University}
\title{\pkg{cgam}: An \proglang{R} Package for the Constrained Generalized Additive Model}

\Plainauthor{Xiyue Liao, Mary C. Meyer}
\Plaintitle{cgam: An R Package for the Constrained Generalized Additive Model} 
\Shorttitle{Constrained Generalized Additive Model}

\Abstract{
The \pkg{cgam} package contains routines to fit the generalized additive model where the components may be modeled with shape and smoothness assumptions. The main routine is {\tt cgam} and nineteen symbolic routines are provided to indicate the relationship between the response and each predictor, which satisfies constraints such as monotonicity, convexity, their combinations, tree, and umbrella orderings.  The user may specify constrained splines to fit the components for continuous predictors, and various types of orderings for the ordinal predictors.  In addition, the user may specify parametrically modeled covariates.  The set over which the likelihood is maximized is a polyhedral convex cone, and a least-squares solution is obtained by projecting the data vector onto the cone. For generalized models, the fit is obtained through iteratively re-weighted cone projections. The cone information criterion is provided and may be used to compare fits for combinations of variables and shapes.  In addition, the routine  {\tt wps} implements monotone regression in two dimensions using warped-plane splines, without an additivity assumption.
The graphical routine {\tt plotpersp} will plot an estimated mean surface for a selected pair of predictors, given an object of either {\tt cgam} or {\tt wps}. This package is now available from the Comprehensive \proglang{R} Archive Network at \url{http://CRAN.R-project.org/package=cgam}. 
}


\Plainkeywords{constrained generalized additive model, isotonic regression, spline regression, partial linear, iteratively re-weighted cone projection, R, graphical routine} 

\Address{
  Xiyue Liao\\
  Department of Statistics\\
  Colorado State University\\
  Fort Collins, Colorado 80523, United States of America\\
  E-mail: \email{xiyue@rams.colostate.edu} \\

  Mary C. Meyer\\
  Department of Statistics\\
  Colorado State University\\
  Fort Collins, Colorado 80523, United States of America\\
  E-mail: \email{meyer@stat.colostate.edu}\\
}

\begin{document}
\newcommand{\bm}[1]{\mbox{\boldmath$#1$}}
\def\R{\mbox{I}\negthinspace \mbox{R}}
\def\hbt{\hat{\bm{\theta}}}
\def\bt{\bm{\theta}}
\def\bp{\bm{\phi}}
\def\hbp{\hat{\bm{\phi}}}
\def\hbe{\hat{\bm{\eta}}}
\def\ve{\varepsilon}
\def\ev{\mbox{E}}
\def\mand{\;\;\mbox{and}\;\;}
\def\mfor{\;\;\mbox{for}\;\;}
\def\C{{\cal C}}

\section[Introduction and overview]{Introduction and overview}  
\label{sec1}
\subsection{Generalized additive model with shape or order constraints}
\label{sec11}
We introduce a comprehensive framework for the generalized additive model with shape or order constraints. 
We consider models with independent observations from an exponential family with density of the form
\begin{equation}
p(y_i;\theta,\tau) = exp[\{y_i\theta_i - b(\theta_i)\}\tau - c(y_i, \tau)],\;\; i = 1,\ldots,n,
\label{eq:model_glm}
\end{equation}
where the specifications of the functions $b$ and $c$ determine the sub-family of models. The mean vector $\bm{\mu} = E(\bm{y})$ has values $\mu_i = b'(\theta_i)$, and is related to a design matrix of predictor variables through a link function $g(\mu_i) = \eta_i$, $i = 1,\ldots,n$.  The link function specifies the relationship with the predictors; for example, suppose $x_1,\ldots,x_L$ are continuous or ordinal predictors and $\bm{z}\in\R^p$ is a vector of covariates to be parametrically modeled.  We specify an additive model  
\begin{equation}
\eta_i = f_1(x_{1i}) + \cdots + f_L(x_{Li}) + \bm{z}^{\top}_i \bm{\alpha},
\label{eq:eta1}
\end{equation}
where the parameter vector $\bm{\alpha} \in \R^p$ and the functions $f_{\ell}$, $\ell=1,\ldots,L$, are to be estimated simultaneously.  The $\eta$ function has been called the ``systematic component'' or the ``linear predictor'' (\citet*{mcc89}, \citet*{has90}); here we will use ``predictor function.''
We consider the Gaussian, Poisson and binomial families in this package; the default is Gaussian.

\begin{figure}
\centering 
\includegraphics[height=2.5in,width=6.1in]{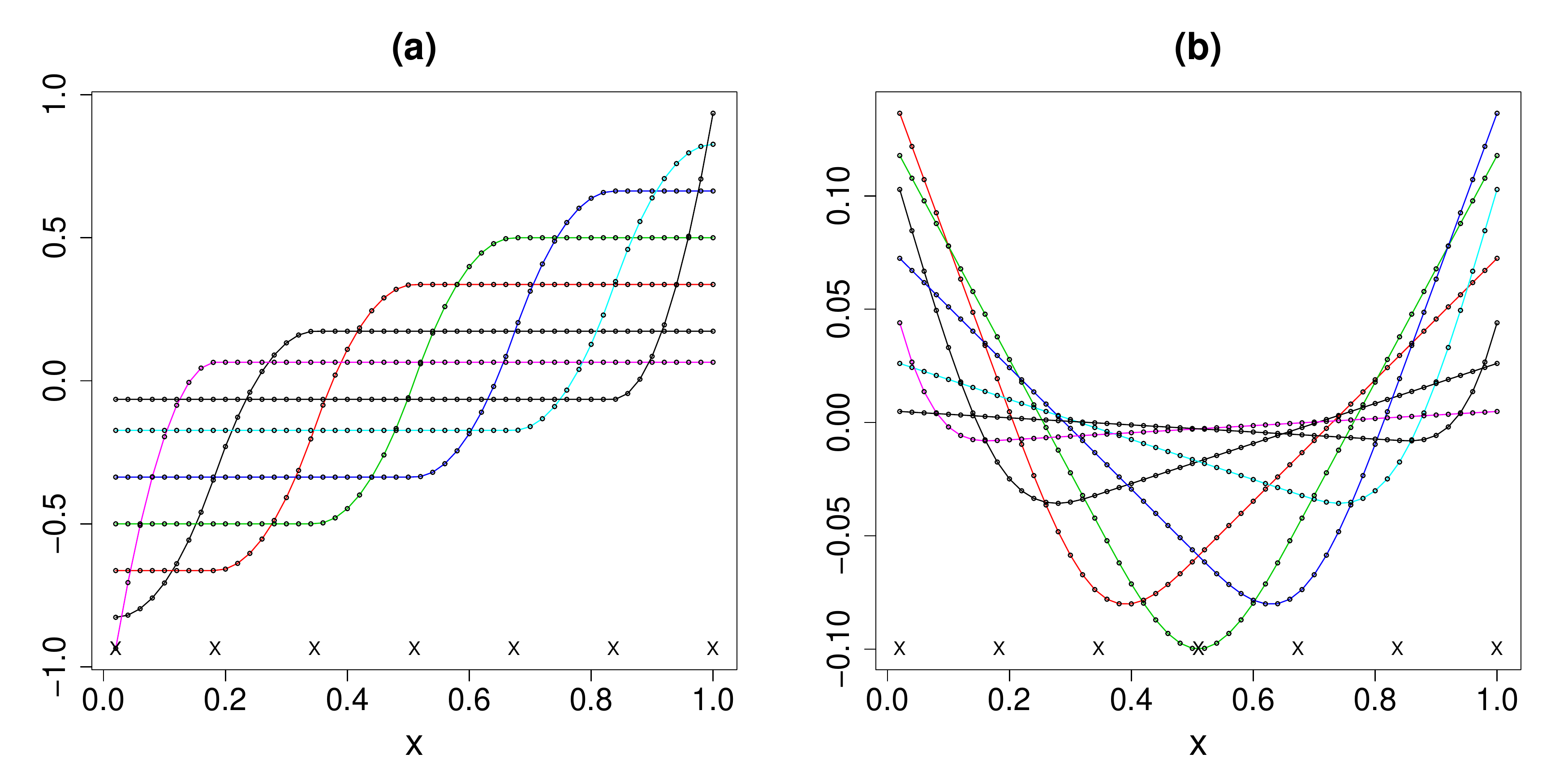}
\captionof{figure}{Smooth spline basis functions using seven equally spaced knots marked as ``X'' for a data set with $n=100$ observations with values marked as dots. (a) smooth and increasing with each basis function centered (b) smooth and convex with each basis function scaled to be orthogonal to $\bm{1}$ and $\bm{x}$.}
\label{basis_sm2}
\end{figure}

  For modeling smooth constrained $f_{\ell}$, there are eight shape choices, i.e., increasing, decreasing, concave, convex, and combinations of monotonicity and convexity. For increasing and decreasing constraints, we use quadratic $I$-spline basis functions, and for constraints involving convexity, cubic $C$-spline basis functions are used.  Example sets of basis functions for seven equally spaced knots are shown in Figure \ref{basis_sm2}; see \citet*{meyer08} for details about these spline bases. 

The $I$-spline basis functions, together with constant function, span the space of piece-wise quadratic splines for the given knots.  The spline function is increasing if and only if the coefficients of the basis functions are positive, and decreasing if and only if the coefficients of the basis functions are negative.    The $C$-spline basis functions, together with the constant function and the identity function, span the space of piece-wise cubic splines for the given knots.  The spline function is convex if and only if the coefficients of the basis functions are positive, and concave if and only if the coefficients of the basis functions are negative.   If we also restrict the sign of the coefficient on the identity function, all four combinations of monotonicity and convexity can be modeled with constrained $C$-splines.   

Define $\bp_{\ell}\in\R^n$ as $\bp_{\ell,i}= f_{\ell}(x_{\ell,i})$, $i=1,\ldots,n$, for a continuous predictor $x_{\ell}$, and define $\bm{s}_{\ell,j}$, $j=1,\ldots,m_{\ell}$ to be the spline basis vectors appropriate for the constraints associated with $x_{\ell}$.  The constraints are satisfied if $\bp_{\ell}\in\C_{\ell}$ where for increasing or decreasing constraints,
\[  \C_{\ell}=\left\{\bp\in\R^n:\bp = a_0\bm{1} + \sum\limits_{j=1}^{m_{\ell}}b_j\bm{s}_j, \;\; b_j\geq 0, \;\; j=1,\ldots, m_{\ell} \right\},
\]
and for convex or concave constraints,
\[  \C_{\ell}=\left\{\bp\in\R^n:\bp= a_0\bm{1}+ a_1\bm{x}+ \sum\limits_{j=1}^{m_{\ell}}b_j\bm{s}_j, \;\; b_j\geq 0, \;\; j=1,\ldots, m_{\ell}\right\}.
\]

For ordinal predictors, constraint cones are defined according to \citet*{meyer13b}.
 Similar to the previous case, there are eight shape constraints involving monotonicity and convexity; in addition, tree and umbrella orderings are options. 
For these orderings, the estimate of $\bm{\phi}_{\ell}$ is in ${\cal C}_{\ell}$, where 
\[{\cal C}_{\ell} = \{\bm{\phi}_{\ell}\in \R^n: \bm{A}_{\ell}\bm{\phi}_{\ell} \geq \bm{0} \hspace{2 mm} \mbox{and} \hspace{2 mm} \bm{B}_{\ell}\bm{\phi}_{\ell} = \bm{0}\},
\]
for constraint matrices $\bm{A}_{\ell}$ and $\bm{B}_{\ell}$ that are $r_{l1} \times n$ and $r_{l2} \times n$, respectively.  The equality constraints handle duplicate values of the predictor, and the rows of $\bm{A}_{\ell}$ and $\bm{B}_{\ell}$ together form a linearly independent set.    

The tree-ordering describes a scenario in which the user assumes that the effect of a categorical variable on the response, for all but one of the levels, is larger (or smaller) than the effect at that level.   This ordering is useful in the case of several treatment levels and a placebo, where it is assumed that the treatment effect is at least that of the placebo, but there are no imposed orderings among the treatments.   For implementation in {\tt cgam}, the level zero is assumed to be the placebo level.

An umbrella ordering is a unimodal assumption on a categorical variable, where the level at which the maximum effect is given.   For implementation in {\tt cgam}, the level zero is used as this mode; other levels are indicated by positive or negative integers.   The effects are ordered on either side of the mode.

See \citet*{meyer13b} for details about construction of basis vectors $\bm{w}_1,\ldots,\bm{w}_{m_{\ell}}$, given $\bm{A}_{\ell}$ and $\bm{B}_{\ell}$, so that
\begin{equation}
{\cal C}_{\ell} = \{\bm{\phi}_{\ell}\in \R^n: \bp_{\ell} = \bm{v}+ \sum_{j=1}^{m_{\ell}} b_j\bm{w}_j, \;\; b_j\geq 0,\;\; j=1,\ldots,m_{\ell}\}.
\label{eq:cone_ls}
\end{equation}
The vector $\bm{v}$ is in a linear space ${\cal V}_{\ell}$ defined by the shape assumptions.   If monotonicity constraints are imposed, ${\cal V}_{\ell}$ is the one-dimensional space of constant vectors; for other types of order constraints, see \citet*{meyer13b} for the construction and composition of ${\cal V}_{\ell}$.

Then $\bm{\eta} = \bp_1+\cdots+\bp_L + \bm{Z}\bm{\beta}$, where $\bp_{\ell}\in\C_{\ell}$ for $\ell=1,\ldots,L$, and the rows of the matrix $\bm{Z}$ are $\bm{z}_i$, $i=1,\ldots,n$.   Each set ${\cal C}_{\ell}$ is a polyhedral convex cone, and let ${\cal V}_z$ be the column space of $\bm{Z}$.   \citet*{meyer13b} showed that $\C = \C_1+\cdots+\C_L+{\cal V}_z$ is also a polyhedral convex cone, where $\bm{\eta}\in\C$ if $\bm{\eta}=\bp_1+\cdots+\bp_L+\bm{v}$ with $\bm{v}\in{\cal V}_z$ and $\bp_{\ell}\in\C_{\ell}$, $\ell=1,\ldots,L$.   That paper also showed how to find a linear space ${\cal L}$ containing the linear spaces ${\cal V}_1,\ldots,{\cal V}_L$ and the column space of $\bm{Z}$, together with ``edge'' vectors $\bm{e}_1,\ldots,\bm{e}_m$ that are orthogonal to ${\cal L}$, so that we can write
\[ \C=\left\{ \bm{\eta}\in\R^n:\bm{\eta}=\bm{v}+\sum_{j=1}^m \alpha_j\bm{e}_j +\bm{Z}\bm{\beta}, \mfor \bm{v}\in{\cal L}, \;\; \alpha_j\geq 0,  \;\; j=1,\ldots,m\right\}.
\]

\subsection{Iteratively re-weighted cone projection}
For the Gaussian family, fitting the additive model (\ref{eq:eta1}) involves a projection of the data vector $\bm{y}$ onto ${\cal C}\subseteq \R^n$.  This is accomplished using the {\tt coneB} routine of the \proglang{R} package \pkg{coneproj} \citet*{xm14}. For binomial and Poisson families, an iteratively re-weighted cone projection algorithm is used. The negative log-likelihood 
\[
L(\theta,\tau;\bm{y}) = \sum\limits_{i=1}^n \Big\{c(y_i, \tau) - \frac{y_i\theta_i - b(\theta_i)}{\tau}\Big\}
\]
is written in terms of the systematic component and minimized over ${\cal C}$. Let $\ell(\bm{\eta})$ be the negative log likelihood written as a function of $\bm{\eta} = (\eta_1,\ldots,\eta_n)^{\top}$. For $\bm{\eta}_k$ in ${\cal C}$, let 
\begin{equation}
\psi_k(\bm{\eta}) = \ell(\bm{\eta}_k) + \nabla \ell(\bm{\eta}_k)^{\top}(\bm{\eta} - \bm{\eta}_k)+\frac{1}{2}(\bm{\eta} - \bm{\eta}_k)^{\top}\bm{Q}_k(\bm{\eta} - \bm{\eta}_k),
\label{eq:ircp}
\end{equation}
where $\nabla \ell(\bm{\eta}_k)$ is the gradient vector and $\bm{Q}_k$ is the Hessian matrix for $\ell(\bm{\eta})$, both evaluated at $\bm{\eta}_k$. The iteratively re-weighted algorithm is:

\begin{enumerate}
\item Choose a valid starting $\bm{\eta}_0$, and set $k = 0$.
\item Given $\bm{\eta}_k$, minimize $\psi_k(\bm{\eta})$ over ${\cal C}$ defined by the model. Then $\bm{\eta}_{k+1}$ minimizes $\ell(\bm{\eta})$ over the line segment connecting the minimizer of $\psi_k(\bm{\eta})$ and $\bm{\eta}_k$.
\item Set $k = k + 1$ and repeat step 2, stopping when a convergence criterion is met.
\end{enumerate}
At step 2, {\tt coneB} is used.  At each iteration of the algorithm, the vector $\bm{\mu}_k$ is computed where $\mu_{ki} = g^{-1}(\eta_{ki})$. If the Hessian matrix is positive definite for all $\bm{\eta}$ then the negative log-likelihood function is strictly convex and $\bm{\mu}_k$ is guaranteed to converge to the MLE $\hat{\bm{\mu}}_k$ (the proof is similar to that in \citet*{meyer04}, Theorem 1).

\subsection{Two-dimensional monotone regression}
\label{sec12}
For two-dimensional isotonic regression without additivity assumptions, the ``warped-plane spline'' (WPS) of \citet*{meyer16b} is implemented in \pkg{cgam} using the function {\tt wps}.  The least-squares model has the form
\begin{equation}
y_i =  f(x_{1i}, x_{2i}) + \bm{z}^{\top}_i \bm{\alpha} +\sigma\ve_i, \mfor i=1,\ldots,n,
\label{eq:model_wps}
\end{equation}
where $\bm{\alpha} \in \R^p$, $\bm{z}_1,\ldots, \bm{z}_n\in\R^p$ contain values of parametrically modeled covariates, and the $\ve_i$'s are mean-zero random errors. We know \emph{a priori} that $f$ is continuous and monotone in both dimensions; that is, for fixed $x_{1}$, if $x_{2a} \leq x_{2b}$, then $f(x_{1}, x_{2a}) \leq f(x_{1}, x_{2a})$, and similarly for fixed $x_{2}$, $f$ is non-decreasing in $x_1$. For linear spline basis functions defined in $x_1$ and $x_2$, the basis functions for the necessary and sufficient constraints are straight-forward and the fitted surface can be described as a continuous piece-wise warped plane. 

\begin{figure}
\centering 
\includegraphics[height=2.2in,width=6.1in]{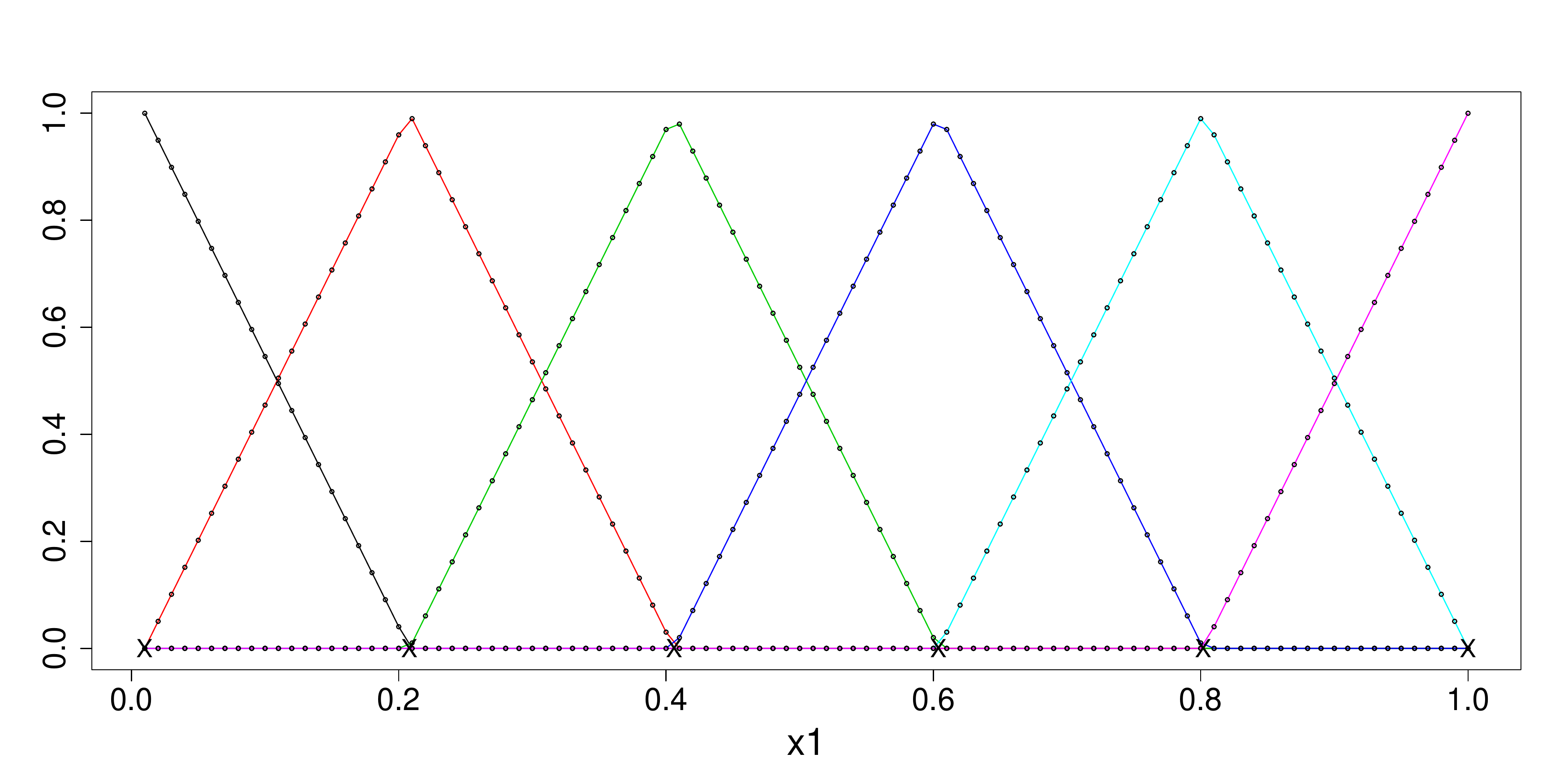}
\captionof{figure}{A set of linear basis functions for a predictor in two-dimensional isotonic regression with $n=100$ observations with values marked as dots and knots marked as ``X''.}
\label{bs_wps}
\end{figure}

Given predictor values $x_{1i}$, $i = 1,\ldots,n$, we define knots $t_{1,1}<\ldots<t_{1,k_1}$, where $t_{1,1}\leq$ min$(\bm{x}_1)$ and $t_{1,k_1} \geq$ max$(\bm{x}_1)$ are defined by evaluating the basis functions at the design points, that is, $\delta_{1,l_1,i} = \delta_{1,l_1}(x_{1,i}), l_1 = 1,\ldots,k_1$. These basis functions span the space of continuous piece-wise linear functions with given knots, and if we replace $\bm{\delta}_{1,1}$ with the constant function $\bm{\delta}_0(\bm{x})=\bm{1}$, then $\bm{\delta}_{0},\bm{\delta}_{1,2},\ldots,\bm{\delta}_{1,k_1}$ span the same space. Similarly, spline basis functions $\bm{\delta}_{2,1},\ldots,\bm{\delta}_{2,k_2}$ can be defined with knots $t_{2,1}<\ldots<t_{2,k_2}$, where $t_{2,1}\leq$ min$(\bm{x}_2)$ and $t_{2,k_2} \geq$ max$(\bm{x}_2)$. An example of a set of basis functions is in Figure \ref{bs_wps}.
Let the $n \times (k_1 - 1)$ matrix $\bm{B}_1$ have as columns $\bm{\delta}_{1,2},\ldots,\bm{\delta}_{1,k_1}$, and let the $n \times (k_2 - 1)$ matrix $\bm{B}_2$ have as columns $\bm{\delta}_{2,2},\ldots,\bm{\delta}_{2,k_2}$. Finally let the $n \times (k_1 - 1)(k_2 - 1)$ matrix $\bm{B}_{12}$ contain the products of basis vectors, so that column $(l_1 - 2)(k_1 - 1) + l_2 - 1$ of $\bm{B}_{12}$ is the element-wise product of $\bm{\delta}_{1,l_1}$ and $\bm{\delta}_{2,l_2}$, for $l_1 = 2,\ldots, k_1$ and $l_2 = 2,\ldots, k_2$. The columns of $\bm{B}_1$, $\bm{B}_2$, and  $\bm{B}_{12}$, together with the vector $\bm{1}$, form a linearly independent set if $n\geq k_1k_2$ and there are no ``empty cells''. 

Let $\bm{\theta}_{ij} = f(x_{1i}, x_{2j})$ be the values of the regression function evaluated at the design points. This is approximated by $\beta_0 \bm{1} + \bm{B}_1\bm{\beta}_1 + \bm{B}_2\bm{\beta}_2  + \bm{B}_{12}\bm{\beta}_3 = \bm{B\beta}$, where $\bm{B} = [\bm{1}|\bm{B}_1|\bm{B}_2|\bm{B}_{12}]$ and $\bm{\beta}^{\top} = [\beta_0|\bm{\beta}_1^{\top}|\bm{\beta}_2^{\top}|\bm{\beta}_3^{\top}]$. A constraint matrix $\bm{A}$ will give the necessary and sufficient conditions, which is stated in {\tt Theorem 1} of \citet*{meyer16b}, for monotonicity of the spline basis functions in both predictors if and only if $\bm{A\beta}\geq \bm{0}$. Here, $\bm{A}$ is a $k \times (k_1 k_2)$ matrix where $k = 2k_1 k_2 - k_1 - k_2$. The constrained least-squares solution is a projection of $\bm{y}$ onto the cone
\begin{equation}
{\cal C} = \{\bm{\mu} \in \R^n: \bm{\mu = B\beta + Z\alpha}; \bm{A\beta} \geq \bm{0}\};
\label{eq:cone_wps}
\end{equation}
the routine {\tt coneA} in the \proglang{R} package \pkg{coneproj} \citet*{xm14} is used.

Penalized warped-plane regression is also included in this package. To obtain smoother fits and to side-step the problem of knot choices, we can use ``large'' $k_1$ and $k_2$, and penalize the changes in slopes of the regression surface, which is a warped plane over each knot rectangle whose slopes can change abruptly from one rectangle to the next.  An additional advantage of penalization is that empty cells are allowed.
The sum of the squared differences in slopes is used as the penalty term, where $\lambda>0$ is a penalty parameter and it will control the constrained ``effective degrees of freedom'' (edfc$_{\lambda}$) of the fit. The standard generalized cross validation (GCV) defined in Chapter 5 of \citet*{rup03} can be used to select a penalty parameter:
\begin{equation}
\mathrm{GCV}(\lambda) = \frac{\sum_{i=1}^{n}[y_i - \hat{\mu}_{\lambda,i}]^2}{(1-\mbox{edfc}_{\lambda}/n)^2}.
\label{eq:gcv}
\end{equation}
A range of values of $\lambda$ can be tried and the GCV choice of $\lambda$ minimizes the criterion. (See \citet*{meyer16b} for more details.)



Inference methods are discussed in Section \ref{sec3}, main routines in this package are discussed in Section \ref{sec4}, and the usage of this package is exemplified with simulated and real data sets in Section \ref{sec5}. In Section \ref{sec6}, we compare this package with some competitor packages in terms of utility and speed.

\section[Inference methods]{Inference methods}  
\label{sec3}
For inference regarding $\bm{\alpha}$, an approximately normal distribution can be derived for $\sqrt{n}(\hat{\bm{\alpha}}-\bm{\alpha})$.  Proposition~4 of \citet*{meyer99} says that there is a subset of edges, indexed by $J\subseteq\{1,\ldots,m\}$ such that the projection of $\bm{y}$ onto $\C$ coincides with projection of $\bm{y}$ onto the linear space spanned by $\bm{e}_j$, $j\in J$, and the basis vectors for ${\cal L}$.  Suppose $\bm{X}_0$ is a matrix so that the columns of $\bm{X}_0$ and $\bm{Z}$ span the linear space ${\cal L}$.   Then if $\bm{P}_J$ is the projection matrix for the spaced spanned by $\bm{e}_j$, $j\in J$ and the columns of $\bm{X}_0$, we can write
\[ \hat{\bm{\alpha}} = [\bm{Z}^{\top}(\bm{I}-\bm{P}_J)\bm{Z}]^{-1}\bm{Z}^{\top}(\bm{I}-\bm{P}_J)\bm{y}.
\]
Under mild regularity conditions, $\hat{\bm{\alpha}}$ is approximately normal with mean zero and covariance $[\bm{Z}^{\top}(\bm{I}-\bm{P}_J)\bm{Z}]^{-1}\sigma^2$.  We estimate $\sigma^2$ as
\[ \hat{\sigma}^2 = \frac{SSR}{n-d_0-cEDF},
\]
where $SSR$ is the sum of squared residuals, $d_0$ is the dimension of the linear space ${\cal L}$, and $EDF$ is the effective degrees of freedom, that is, the cardinality of $J$ plus the dimension of ${\cal L}$.   The constant $c$ is between 1 and 2;  \citet*{meyer00} showed this multiplier is appropriate for cone regression.    That paper gave evidence that $c=1.5$ is appropriate for unsmoothed isotonic regression; simulations suggest that for constrained splines, a smaller value gives better estimates of $\sigma^2$. In the {\tt cgam} routine and the {\tt wps} routine, the default is  $c=1.2$, but the user can specify $c\in[1,2]$. 

These results are used to construct approximate $t$ and $F$~tests for $\bm{\alpha}$; specifically, $\hat{\bm{\alpha}}$ is taken to be approximately normal with mean $\bm{\alpha}$ and covariance  $[\bm{Z}^{\top}(\bm{I}-\bm{P}_J)\bm{Z}]^{-1}\hat{\sigma}^2$.   See \citet*{meyer16a} and \citet*{meyer16b} for detailed conditions under which this is a good approximation.


The cone information criterion (CIC) proposed in \citet*{meyer13} for the least-squares model is generalized to
\[  CIC = -\frac{2}{n}log(L) + log\left\{\frac{2\big[E_0(EDF) + d_0\big]}{n-d_0-1.5E_0(EDF)}+ 1\right\},
\]
where $L$ is the likelihood maximized over the cone ${\cal C}$, $d_0$ is the dimension of the linear space ${\cal L}$, and $E_0(EDF)$ is the null expected dimension of the face of ${\cal C}$ on which the projection lands.   To compute $E_0(EDF)$, we simulate from~(\ref{eq:model_glm}) and~(\ref{eq:eta1}) with $f_{\ell}\equiv 0$ for $\ell=1.\ldots,L$.  In this way we get the expected degrees of freedom for the constrained model, in the case where the $f_{\ell}$ do not contribute to the expected response.  This is appropriate for model selection, as the observed $EDF$ tends to be larger when the predictors are related to the response.  See \citet*{meyer13} for more details about using the CIC for model selection.

This criterion is the estimated predictive squared error, similar to the AIC, and is specially derived for cone projection problems. If the constraints are not known \emph{a priori}, the CIC model selection procedure may be used to select not only the variables in a model of the form (\ref{eq:eta1}), but also the nature of their relationships with the response.

\section[Main routines in this package]{Main routines in this package}  
\label{sec4}
The function {\tt cgam} is the main routine which implements the constrained generalized additive regression. For a non-parametrically modeled effect, a shape restriction can be imposed on the predictor function component with optional smoothing, or a partial ordering can be imposed. An arbitrary number of parametrically modeled covariates may be included. The user can also choose an unconstrained smooth fit for one or more of the $f_{\ell}$, which is simply the least-squares estimator using the set of cubic spline basis functions created for convex constraints.  
The specification of the model in {\tt cgam} uses one or more of nineteen symbol functions to specify the shape, ordering, and smoothness of each $f_{\ell}$.

\subsection[The symbolic functions that specify the form of f]{The symbolic functions that specify the form of $f_{\ell}$}
\label{sec40}
To specify an effect that is increasing with a predictor $x$ without smoothing, the function {\tt incr(x)} is used in the statement of the {\tt cgam} routine.  Other functions for unsmoothed effects are {\tt decr}, {\tt conv}, {\tt conc}, {\tt incr.conv}, {\tt incr.conc}, {\tt decr.conv}, {\tt decr.conc}, {\tt tree}, and {\tt umbrella}.

For smooth estimates of $f_{\ell}$, the following functions may be used: {\tt s.incr}, {\tt s.decr}, {\tt s.conv}, {\tt s.conc}, {\tt s.incr.conv}, {\tt s.incr.conc}, {\tt s.decr.conv}, and {\tt s.decr.conc}.   For fitting an unconstrained smooth effect, {\tt s(x)} may be used.   Each of these nineteen functions implements a routine to create the appropriate  set of basis functions.   The smoothed versions have options for number and spacing of knots.   For example, {\tt s.decr(x, numknots = 10, space = "Q")} will create I-spline basis functions with ten knots at equal quantiles of the observed {\tt x} values.  The default is {\tt space = "E"} which provides equal spacing. For a data set of $n$ observations, the number of knots has a default of order $n^{1/9}$ ($n^{1/7}$) when C-spline (I-spline) basis functions are used.



\subsection[Basic usage of the main routine: cgam]{Basic usage of the main routine: {\tt cgam}}
\label{sec41}

In the {\tt cgam} routine, the specification of the model can be one of the three exponential families: Gaussian, Poisson and binomial. The symbolic functions are used to specify how the predictors are related to the response. For example,
\begin{Schunk}
\begin{Sinput}
R> fit <- cgam(y ~ s.incr.conv(x1) + s(x2, numknots = 5), family = gaussian())
\end{Sinput}
\end{Schunk}
 specifies that the response $\bm{y}$ is from the Gaussian distribution, and $E(\bm{y})$ is smoothly increasing and concave in {\tt x1}, while component function for {\tt x2} is smooth but unconstrained, with five equally spaced knots. 

The user can also specify the parameter {\tt nsim} to simulate the CIC value of the model. Such simulation can be time-consuming. The default is {\tt nsim} = 100. For example, we can write 
\begin{Schunk}
\begin{Sinput}
R> fit <- cgam(y ~ s.incr.conv(x1, numknots = 10, space = "Q") 
+  s(x2, numknots = 10, space = "Q"), family = gaussian(), nsim = 1000)
\end{Sinput}
\end{Schunk}
For a {\tt cgam} fit, the main values returned are the estimated systematic component $\hat{\bm{\eta}}$ and the estimated mean value $\hat{\bm{\mu}}$, obtained by transforming $\hat{\bm{\eta}}$ by the inverse of the link function. The CIC value will also be returned if the user chooses to simulate it.

The routine {\tt summary} provides the estimates, the standard errors, and approximate $t$~values and $p$~values for the linear terms. A {\tt summary} table also includes the deviance for the null model of a {\tt cgam} fit, i.e., the model only containing the constant vector and the residual deviance of a {\tt cgam} fit.

\subsection[Basic usage of the main routine: wps]{Basic usage of the main routine: {\tt wps}}
\label{sec42}
The {\tt wps} routine implements two-dimensional isotonic regression using warped-plane splines discussed in \citet*{meyer16b}. Parametrically-modeled covariates can be incorporated in the regression. 
Three symbolic subroutines are used in a {\tt wps} formula to specify the relationship between the mean surface and the two predictors to be one of the three: ``doubly-decreasing'', ``doubly-increasing'', and ``decreasing-increasing''. To be specific, a basic {\tt wps} formula will look like 
\begin{Schunk}
\begin{Sinput}
R> fit.dd <- wps(y ~ dd(x1, x2))
\end{Sinput}
\end{Schunk}
The argument {\tt dd} specifies that $E(\bm{y})$ is decreasing in both $\bm{x}_1$ and $\bm{x}_2$. Similarly, {\tt ii} specifies that the relationship is doubly-increasing, and {\tt di} specifies that  $E(\bm{y})$ is decreasing in $\bm{x}_1$ and increasing in $\bm{x}_2$. The knots vector for each predictor can be defined similarly as a {\tt cgam} fit. 

The user can also choose to use a penalized version by providing a ``large'' number of knots for $\bm{x}_1$ and $\bm{x}_2$ with a penalty term. For example, we want to use ten equally spaced knots for each predictor with a penalty term to be $.1$ in a ``doubly-decreasing'' formula. Then we can write 
\begin{Schunk}
\begin{Sinput}
R> fit.dd <- wps(y ~ dd(x1, x2, numknots = c(10, 10), space = c("E", "E")),
+  pen = .1)
\end{Sinput}
\end{Schunk}
For a {\tt wps} fit, the main values returned are the estimated mean value $\hat{\bm{\mu}}$ and the constrained effective degrees of freedom (edfc). The generalized cross validation value (GCV) for the constrained fit is also returned, which could be used to choose the penalty parameter $\lambda$. The {\tt summary} routine works the same way as for a {\tt cgam} fit when there is any parametrically modeled covariate.

\subsection[Basic usage of the graphical routine: plotpersp]{Basic usage of the graphical routine: {\tt plotpersp}}
\label{sec43}
This routine is an extension of the generic \proglang{R} graphics routine {\tt persp}. For a {\tt cgam} object, which has at least two non-parametrically modeled predictors, this routine will make a three-dimensional plot of the fit with a set of two non-parametrically modeled predictors in the formula, which will be marked as the {\tt x} and {\tt y} labs in the plot. If there are more than two non-parametrically modeled predictors, any other such predictor will be evaluated at the largest value which is smaller than or equal to its median value. This routine also works for a {\tt wps} object, which only has two isotonically modeled predictors. For a {\tt cgam} fit, the {\tt z} lab represents the estimated regression surface of the mean or the systematic component according to the user's choice, and for a {\tt wps} fit, the {\tt z} lab represents the constrained or the unconstrained regression surface of the mean according to the user's choice. If there is any categorical covariate in a {\tt cgam} or {\tt wps} model and if the user specifies the argument {\tt categ} to be a character representing a categorical covariate in the formula, then a three-dimensional plot with multiple parallel surfaces, which represent the levels of the categorical covariate in an ascending order, will be created; otherwise, a three-dimensional plot with only one surface will be created. Each level of a categorical covariate will be evaluated at its mode. 

The basic form of this routine is defined as
\begin{Schunk}
\begin{Sinput}
R> plotpersp(object,...)
\end{Sinput}
\end{Schunk}
The argument {\tt object} represents an object of the {\tt wps} class, or an object of the {\tt cgam} class with at least two non-parametrically modeled predictors. When there are more than two non-parametrically modeled predictors in a {\tt cgam} formula, the user may choose to write 
\begin{Schunk}
\begin{Sinput}
R> plotpersp(object, x1, x2,...)
\end{Sinput}
\end{Schunk}
The arguments {\tt x1} and {\tt x2} represent two non-parametrically modeled predictors in the model. If the user omits the two arguments, the first two non-parametrically modeled predictors in the formula will be used.
 
\section[User guide]{User guide}  
\label{sec5}
We demonstrate the main routines using simulated data sets and real data sets in detailed examples. For a more complete explanation and demonstration of each routine, see the official reference manual of this package at \url{http://CRAN.R-project.org/package=cgam}. This package depends on the package \pkg{coneproj}, which conducts the cone projection algorithm. (See \citet*{xm14} for more details.)
\begin{Schunk}
\begin{Sinput}
R> install.packages("cgam")
R> library("cgam")
Loading required package: coneproj
\end{Sinput}
\end{Schunk}
\subsection[Fitting constrained surface to "Rubber"]{Fitting constrained surface to ``Rubber''}  
\label{sec51}
\begin{figure}
\centering 
\includegraphics[height=3.0in,width=6.3in]{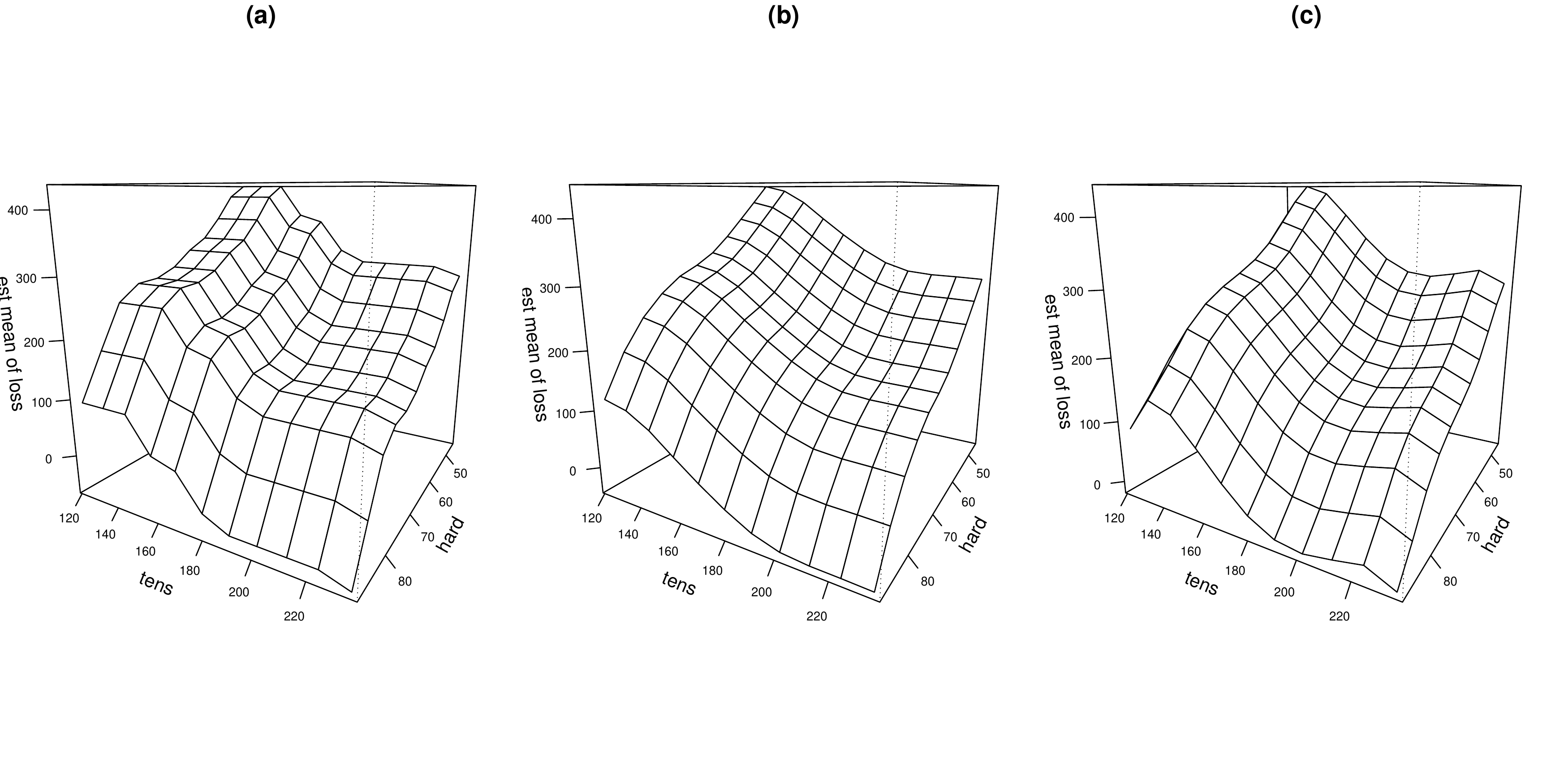}
\vspace{-2mm}
\captionof{figure}{Demonstration of constrained regression using the {\tt cgam} routine with the ``Rubber'' data set. (a) the estimated surface is decreasing in both predictors without smoothing. (b) the estimated surface is smooth and decreasing in both predictors. (c) the estimated surface is smooth in both predictors without shape constraint.}
\label{fig:rubber}
\end{figure}

The ``Rubber'' data set in the package \pkg{MASS} has $30$ observations and three variables relevant to accelerated testing of tyre rubber, i.e.,  ``loss'' (abrasion loss), ``hard'' (hardness), and ``tens'' (tensile strength). Assuming that ``loss'' is decreasing in both ``hard'' and ``tens'', the effects are additive, and the response is Gaussian, we can model the relationship as following:

\begin{Schunk}
\begin{Sinput}
R> library("MASS") 
R> data("Rubber")
R> loss <- Rubber$loss
R> hard <- Rubber$hard
R> tens <- Rubber$tens
R> set.seed(123)
R> fit.decr <- cgam(loss ~ decr(hard) + decr(tens))
\end{Sinput}
\end{Schunk}
Alternatively, we can model the relationship using $I$-spline basis functions:
\begin{Schunk}
\begin{Sinput}
R> set.seed(123)
R> fit.s.decr <- cgam(loss ~ s.decr(hard) + s.decr(tens))
\end{Sinput}
\end{Schunk}
For a spline-based fit without constraints:
\begin{Schunk}
\begin{Sinput}
R> set.seed(123)
R> fit.s <- cgam(loss ~ s(hard) + s(tens))
\end{Sinput}
\end{Schunk}
For each fit, we use the default that {\tt nsim = 100} to get the CIC parameter. According to the CIC value of each fit, the non-smooth fit is better than another two smooth fits for this data set.
\begin{Schunk}
\begin{Sinput}
R> fit.decr$cic
[1] 8.945583
R> fit.s.decr$cic
[1] 10.05648
R> fit.s$cic
[1] 10.16878
\end{Sinput}
\end{Schunk}
We can call the routine {\tt plotpersp} to make a 3D plot of the estimated mean surface based on {\tt fit.decr}, {\tt fit.s.decr}, and {\tt fit.s}, which is shown in Figure \ref{fig:rubber}.
\begin{Schunk}
\begin{Sinput}
R> par(mfrow = c(1, 3))
R> plotpersp(fit.decr, hard, tens, th = 120, main = "(a)")
R> plotpersp(fit.s.decr, hard, tens, th = 120, main = "(b)")
R> plotpersp(fit.s, hard, tens, th = 120, main = "(c)")
\end{Sinput}
\end{Schunk}

\subsection[Fitting parallel surfaces to "plasma"]{Fitting parallel surfaces to ``plasma''}  
\label{sec52}
The ``plasma'' data set, available at \url{http://axon.cs.byu.edu/data/statlib/numeric/plasma_retinol.arff}, contains $314$ observations of blood plasma beta carotene measurements along with several covariates. High levels of blood plasma beta carotene are believed to be protective against cancer, and it is of interest to determine the relationships with covariates. Here we use the logarithm of  ``plasma'' level as the response, and choose ``bmi'',  the logarithm of  ``dietfat'', ``cholest'', ``fiber'',  ``betacaro'' and ``retinol'' as shape-restricted predictors. In addition, we include ``smoke'' and ``vituse'' as categorical covariates.

\begin{Schunk}
\begin{Sinput}
R> set.seed(123)
R> fit <- cgam(logplasma ~ s.decr(bmi) + s.decr(logdietfat) + s.decr(cholest) 
+  + s.incr(fiber) + s.incr(betacaro) + s.incr(retinol) + factor(smoke) 
+  + factor(vituse), data = plasma)
\end{Sinput}
\end{Schunk}

\begin{figure}
\centering 
\includegraphics[height=3.0in,width=6.3in]{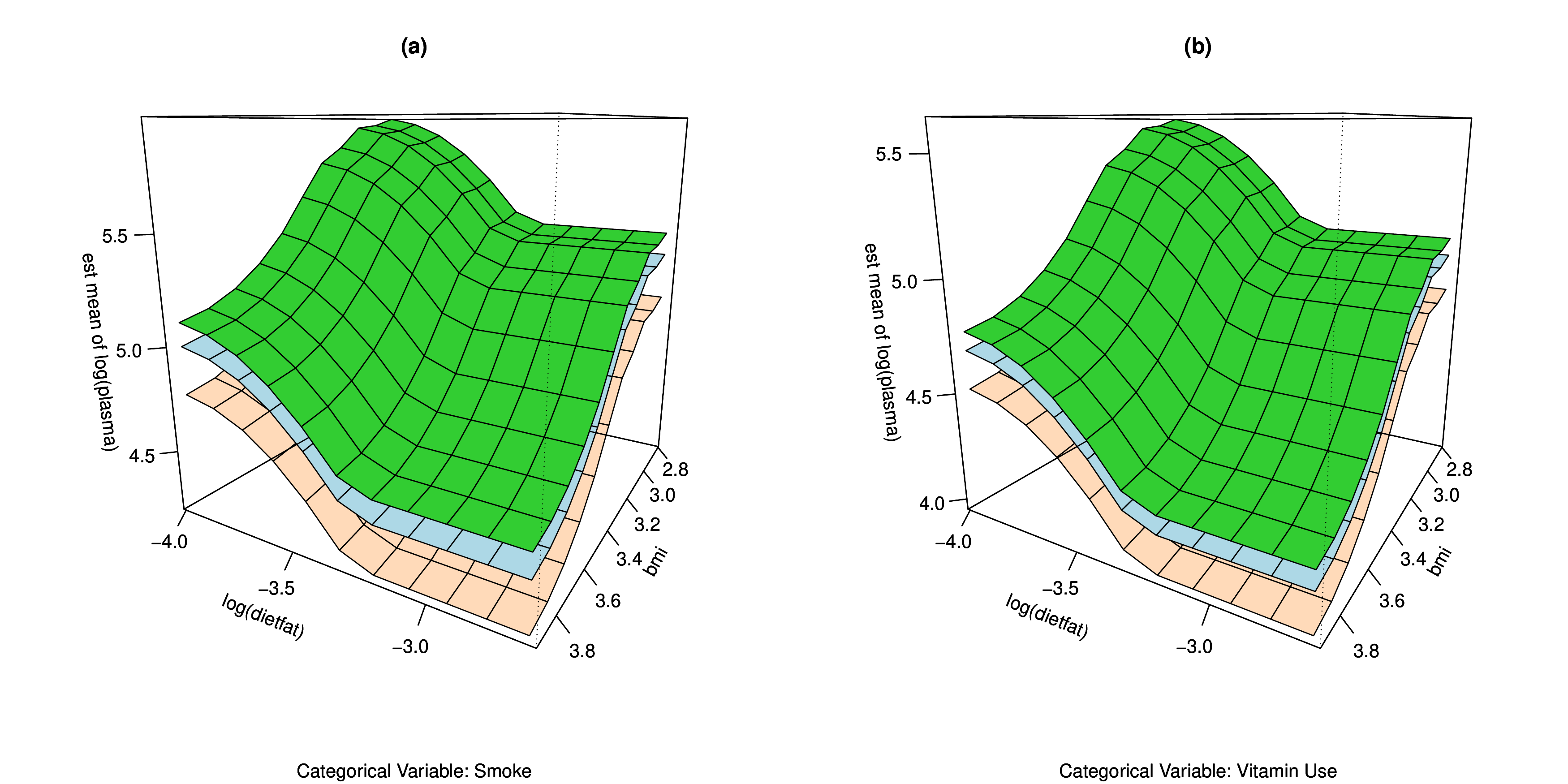}
\captionof{figure}{Demonstration of constrained regression using the {\tt cgam} function with the ``plasma'' data set. (a) parallel surfaces representing the effects of three levels of ``smoke'' in an ascending order. (b) parallel surfaces representing the effects of three levels of ``vituse'' in an ascending order.}
\label{fig:plasma}
\end{figure}

We can call  {\tt summary} to check the estimate, the standard error, the approximate $t$~value and the $p$~value for the coefficient of the categorical covariates. The CIC value is also simulated and returned.
\begin{Schunk}
\begin{Sinput}
R> summary(fit)
\end{Sinput}
\begin{Soutput}
Call:
cgam(formula = logplasma ~ s.decr(bmi) + s.decr(logdietfat) + s.decr(cholest)
  + s.incr(fiber) + s.incr(betacaro) + s.incr(retinol) + factor(smoke) 
  + factor(vituse), data = plasma)

Coefficients:
                Estimate  StdErr t.value p.value    
(Intercept)       4.8398  0.1298 37.2895  <2e-16 ***
factor(smoke)2    0.2145  0.1279  1.6769  0.0947 .  
factor(smoke)3    0.3232  0.1272  2.5402  0.0116 *  
factor(vituse)2  -0.0936  0.1032 -0.9070  0.3652    
factor(vituse)3  -0.2688  0.0948 -2.8345  0.0049 ** 
---
Signif. codes:  0 '***' 0.001 '**' 0.01 '*' 0.05 '.' 0.1 ' ' 1

(Dispersion parameter for gaussian family taken to be 0.4164 )

Null deviance: 174.9801 on 313 degrees of freedom
Residual deviance: 128.6684 on 277.8 observed degrees of freedom
CIC: 4.9661
\end{Soutput}
\end{Schunk}
Again, we use {\tt plotpersp} to show the estimated mean surface based on the fit in Figure \ref{fig:plasma}, where {\tt xlab} is ``bmi'', {\tt ylab} is  the logarithm of ``dietfat'' , and the effects of the levels of each categorical covariate are shown in an ascending order. Other shape-restricted predictors  are evaluated at the median value. 
\subsection[Partial-Ordering examples: tree-ordering and umbrella-ordering]{Partial-Ordering examples: tree-ordering and umbrella-ordering}  
\label{sec53}
\begin{figure}
\centering 
\includegraphics[height=2.8in,width=6.2in]{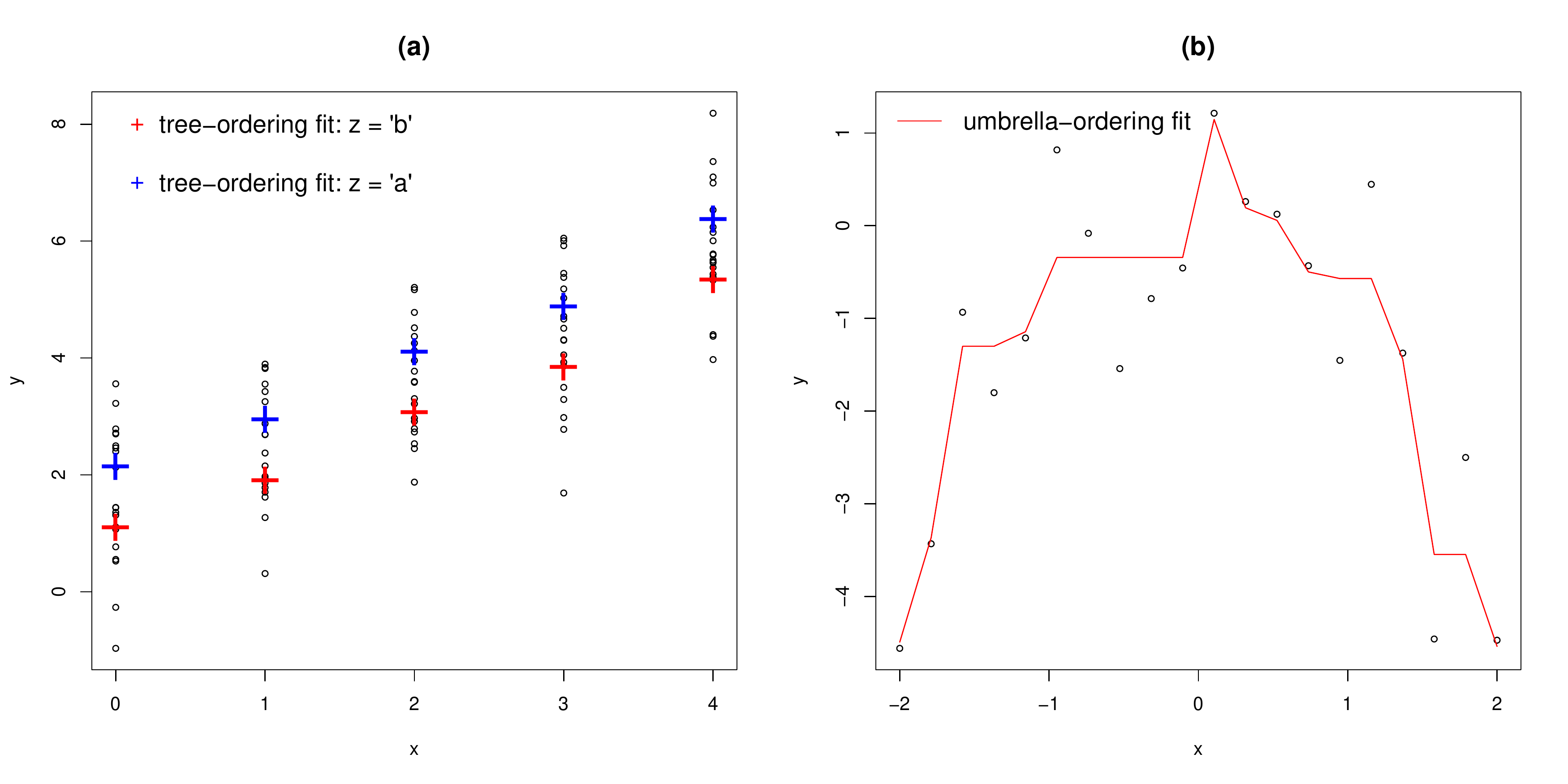}
\captionof{figure}{Demonstration of constrained regression using the {\tt cgam} function with a partial ordering constraint. (a) a tree-ordering fit with a categorical covariate $\bm{z}$. (b) an umbrella-ordering fit.}
\label{fig:ord}
\end{figure}

We simulate a data set as a tree-ordering example such that $\bm{x}$ is a categorical variable with five levels: $x_1=0$ (placebo), $x_2=1$, $x_3=2$, $x_4=3$ and $x_5=4$. Each level has $20$ observations. We also include a categorical covariate $\bm{z}$ with two levels ``a'' and ``b'', which could be a treatment variable people are concerned about, in the model such that when $\bm{x}$ is fixed, the mean response is one unit larger if $\bm{z}$ is ``a''. We use {\tt cgam} to estimate the effect for each level given $\bm{z}$. The fit is in Figure \ref{fig:ord}(a).
\begin{Schunk}
\begin{Sinput}
R> n <- 100 
R> x <- rep(0:4, each = 20)
R> z <- rep(c("a", "b"), 50)
R> y <- x + I(z == "a") + rnorm(n, 1)
R> fit.tree <- cgam(y ~ tree(x) + factor(z)) 
\end{Sinput}
\end{Schunk}
The estimated effect of $\bm{z}$ can be checked by {\tt summary}.
\begin{Schunk}
\begin{Sinput}
R> summary(fit.tree)
Coefficients:
            Estimate  StdErr t.value   p.value    
(Intercept)   2.1617  0.2292  9.4324 < 2.2e-16 ***
factor(z)b   -1.0402  0.1871 -5.5588 < 2.2e-16 ***
---
Signif. codes:  0 '***' 0.001 '**' 0.01 '*' 0.05 '.' 0.1 ' ' 1
\end{Sinput}
\end{Schunk}
For an umbrella-ordering example, we simulate a data set such that $x_0=0$ (mode) and for $x_1, x_2 \geq x_0$, the estimated mean curve is decreasing, while for $x_1, x_2 \leq x_0$, it is increasing. The fit is in Figure \ref{fig:ord}(b).
\begin{Schunk}
\begin{Sinput}
R> n <- 20
R> x <- seq(-2, 2, length = n)
R> y <- - x^2 + rnorm(n)
R> fit.umb <- cgam(y ~ umbrella(x)) 
\end{Sinput}
\end{Schunk}

\subsection[Binomial response example]{Binomial response example}  
\label{sec54}
\begin{figure}
\centering 
\includegraphics[height=3.0in,width=6.3in]{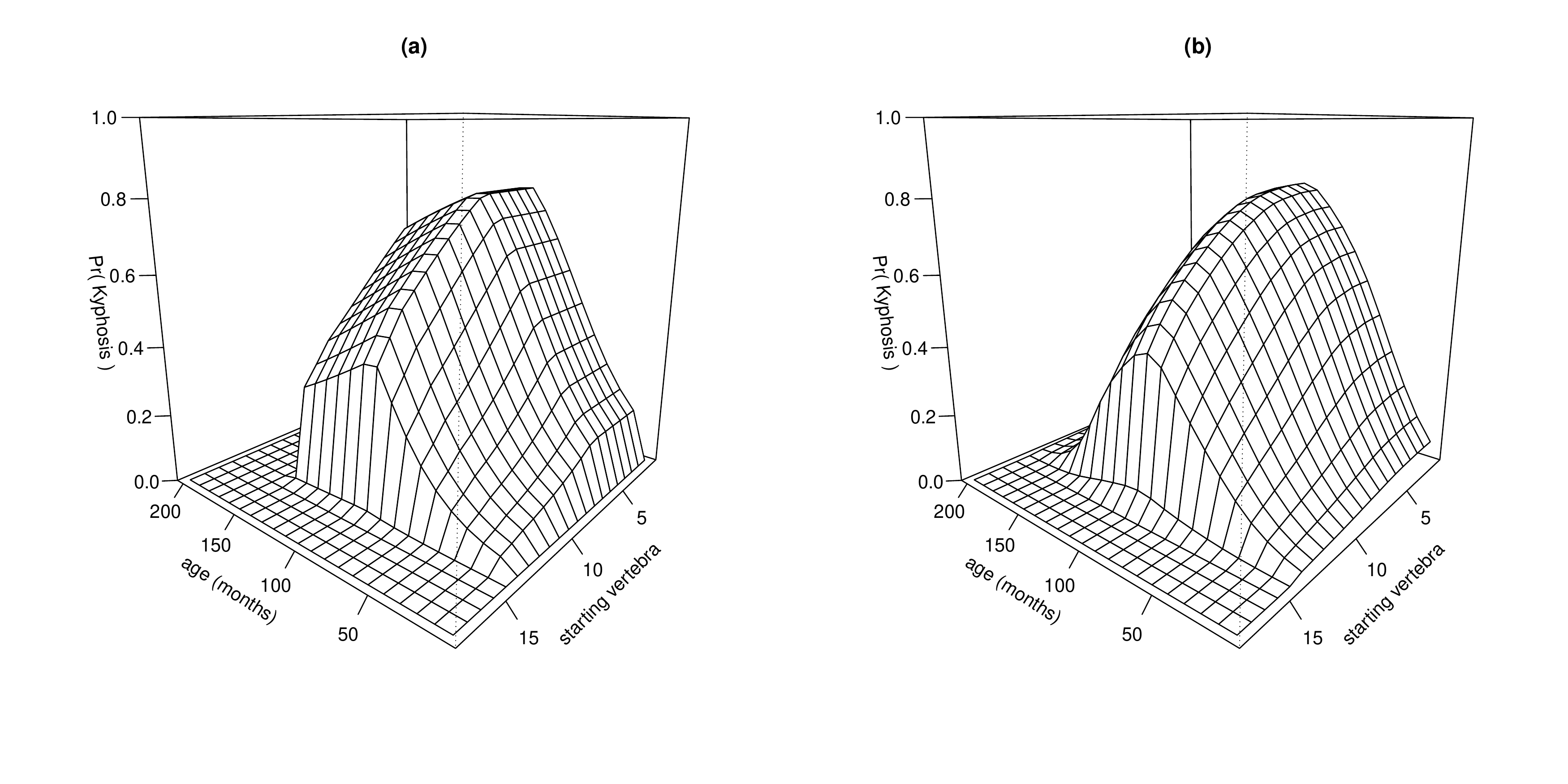}
\captionof{figure}{Demonstration of constrained regression using the {\tt cgam} function with the ``kyphosis'' data set. The surface represents the estimated probability of the response Kyphosis to be present. (a) surface without smoothing (b) smooth surface.}
\label{fig:kypho}
\end{figure}

We use the ``kyphosis'' data set with $81$ observations in the \pkg{gam} package to show that how {\tt cgam} works given a binomial response. In this example, we treat the variable ``Kyphosis'' as the response which is binary, and model the log-odds of ``Kyphosis'' as concave in ``Age'' (age of child in months), increasing-concave in ``Number'' (number of vertebra involved in the operation), and decreasing-concave in ``Start'' (start level of the operation). The non-smooth fit and the smooth fit are shown in Figure \ref{fig:kypho} by {\tt plotpersp}.
\begin{Schunk}
\begin{Sinput}
R> library("gam")
R> data("kyphosis")
R> fit <- cgam(Kyphosis ~ conc(Age) + incr.conc(Number) + decr.conc(Start), 
+  family = binomial(), data = kyphosis)  
R> fit.s <- cgam(Kyphosis ~ s.conc(Age) + s.incr.conc(Number)
+  + s.decr.conc(Start), family = binomial(), data = kyphosis) 
\end{Sinput}
\end{Schunk}

\begin{figure}
\centering 
\includegraphics[height=3.0in,width=6.3in]{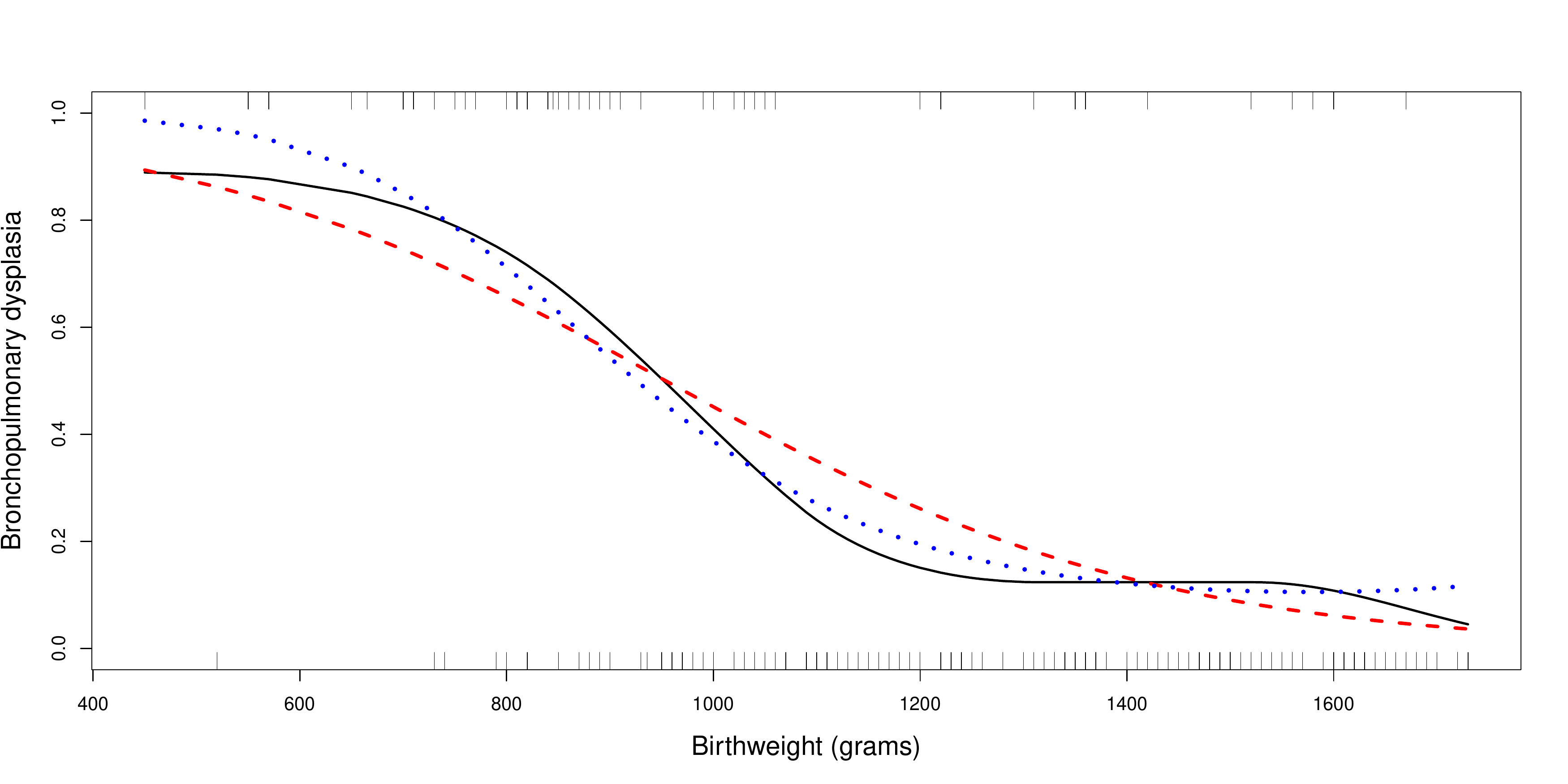}
\captionof{figure}{The estimated probability of bronchopulmonary dysplasia as a function of birth weight. The data are shown as tick marks at the presence (1) and the absence (0) of the condition. The solid curve is the smoothly decreasing fit, the dashed curve is the linear log-odds fit, and the dotted curve is the quadratic log-odds fit.}
\label{fig:bpd}
\end{figure}

Next, we consider the ``bpd'' data set in the \pkg{SemiPar} package. It has $223$ observations with two variables: birth weight of babies and BPD, which is a binary variable indicating the presence of bronchopulmonary dysplasia. It is known that bronchopulmonary dysplasia is more often found in babies with low birth weight, and we can model the relationship between the probability of bronchopulmonary dysplasia and birth weight as smoothly decreasing. The fit is shown in Figure \ref{fig:bpd}. We also include the linear and quadratic log-odds fit in the plot as a comparison. The linear log-odds fit might overly simplify the underlying relationship, while the quadratic fit starts increasing at the end although it seems to be better than the linear fit. 

\begin{Schunk}
\begin{Sinput}
R> library("SemiPar")
R> data("bpd")
R> BPD <- bpd$BPD
R> birthweight <- bpd$birthweight
R> fit.s.decr <- cgam(BPD ~ s.decr(birthweight, space = "Q"), 
+  family = binomial())
\end{Sinput}
\end{Schunk}

\subsection[Poisson response example]{Poisson response example}  
\label{sec55}
\begin{figure}
\centering 
\includegraphics[height=3.0in,width=6.3in]{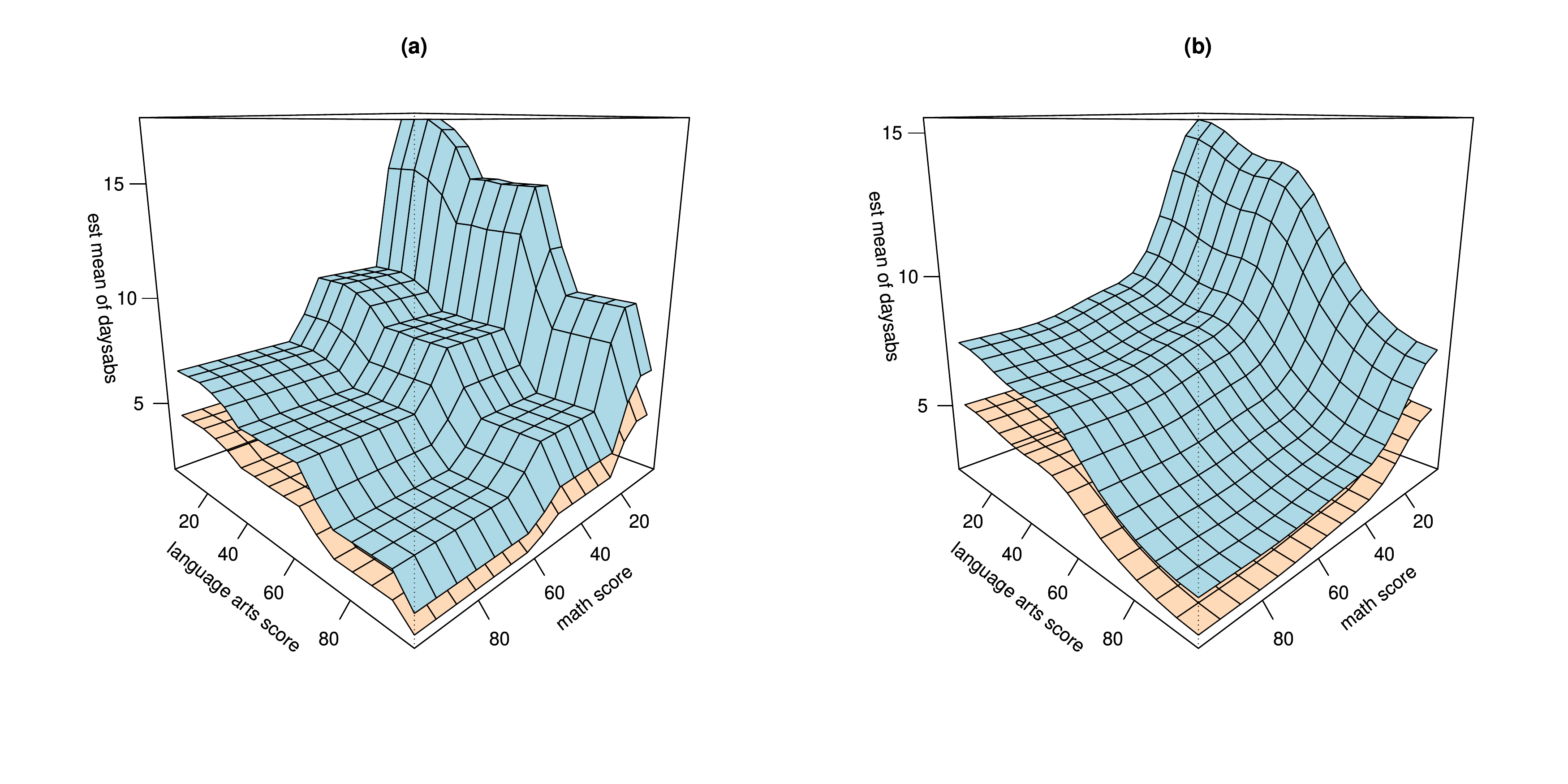}
\captionof{figure}{Demonstration of constrained regression using the {\tt cgam} function with the ``attendance'' data set. (a) parallel surfaces representing the gender effect in an ascending order. (b) parallel surfaces representing the gender effect in an ascending order with smoothing.}
\label{fig:pois}
\end{figure}

Another data set is an attendance data set of $316$ high school juniors from two urban high schools. We use the variable ``daysabs'' (days absent) as a Poisson response. The variables ``math'' and ``langarts'' are the standardized test scores for math and language arts. A categorical variable ``male'' is also included in this data set, which indicates the gender of a student. With \emph{a priori} knowledge that ``daysabs'' is decreasing with respect to each continuous predictor, we can try modeling the relationship between  ``daysabs'' and ``math'' and ``langarts'' as decreasing with ``male'' as a categorical covariate. First, we model the relationship with ordinal basis functions.
\begin{Schunk}
\begin{Sinput}
R> set.seed(123)
R> fit.cgam <- cgam(daysabs ~ decr(math) + decr(langarts) + factor(male), 
+  family = poisson())
R> summary(fit.cgam)
Call:
cgam(formula = daysabs ~ decr(math) + decr(langarts) + factor(male), 
  family = poisson())

Coefficients:
              Estimate  StdErr z.value   p.value    
(Intercept)     1.8715  0.0317 58.9843 < 2.2e-16 ***
factor(male)1  -0.4025  0.0496 -8.1155 < 2.2e-16 ***
---
Signif. codes:  0 '***' 0.001 '**' 0.01 '*' 0.05 '.' 0.1 ' ' 1

(Dispersion parameter for poisson family taken to be 1)

Null deviance: 2409.8204 on 315 degrees of freedom
Residual deviance: 2100.6701 on 296 observed degrees of freedom
CIC: -9.7546
\end{Sinput}
\end{Schunk}
Next, we try modeling the relationship with smooth $I$-splines. 
\begin{Schunk}
\begin{Sinput}
R> set.seed(123)
R> fit.cgam.s <- cgam(daysabs ~ s.decr(math) + s.decr(langarts) + factor(male), 
+  family = poisson())
R> summary(fit.cgam.s)
Call:
cgam(formula = daysabs ~ s.decr(math) + s.decr(langarts) + factor(male), 
  family = poisson())
    
Coefficients:
              Estimate  StdErr z.value   p.value    
(Intercept)     1.8982  0.0309 61.4593 < 2.2e-16 ***
factor(male)1  -0.3988  0.0487 -8.1894 < 2.2e-16 ***
---
Signif. codes:  0 '***' 0.001 '**' 0.01 '*' 0.05 '.' 0.1 ' ' 1

(Dispersion parameter for poisson family taken to be 1)

Null deviance: 2409.8204 on 315 degrees of freedom
Residual deviance: 2201.237 on 304.4 observed degrees of freedom
CIC: -9.4567
\end{Sinput}
\end{Schunk}
According to the simulated CIC value of each fit, it is suggested that the non-smooth fit is better than the fit using smooth $I$-splines. Moreover, the gender effect is significant in both fits. The fits are shown in Figure \ref{fig:pois}.

\subsection[Fitting "doubly-decreasing" surface to "plasma"]{Fitting ``doubly-decreasing'' surface to ``plasma''} 
\label{sec56}
We again use the ``plasma'' data set as an example to illustrate the routine {\tt wps}, and now we assume that the logarithm of ``plasma'' is doubly-decreasing in ``bmi'' and the logarithm of ``dietfat'', and the effects of the two predictors are not necessarily additive. We also include ``smoke'' and ``vituse'' as two categorical covariates. We can model the relationship as following, and we choose $10$ equally-spaced knots for each predictor with the penalty term to be $.505$, which is calculated inside {\tt wps} and can be checked as an output:
\begin{Schunk}
\begin{Sinput}
R> fit <- wps(logplasma ~ dd(bmi, logdietfat) + factor(smoke) + factor(vituse), 
+  data = plasma, pnt = TRUE) 
R> fit$pen
[1] 0.5047263
R> summary(fit)
Call:
wps(formula = logplasma ~ dd(bmi, logdietfat) + factor(smoke) + factor(vituse), 
  data = plasma, pnt = TRUE)

Coefficients:
                Estimate  StdErr t.value p.value    
(Intercept)       4.0144  0.1248 32.1730  <2e-16 ***
factor(smoke)2    0.2988  0.1261  2.3689  0.0185 *  
factor(smoke)3    0.4184  0.1228  3.4079  0.0007 ***
factor(vituse)2  -0.0667  0.1013 -0.6581  0.5110    
factor(vituse)3  -0.2757  0.0935 -2.9490  0.0034 ** 
---
Signif. codes:  0 '***' 0.001 '**' 0.01 '*' 0.05 '.' 0.1 ' ' 1
\end{Sinput}
\end{Schunk}

\begin{figure}
\centering 
\includegraphics[height=3.0in,width=6.5in]{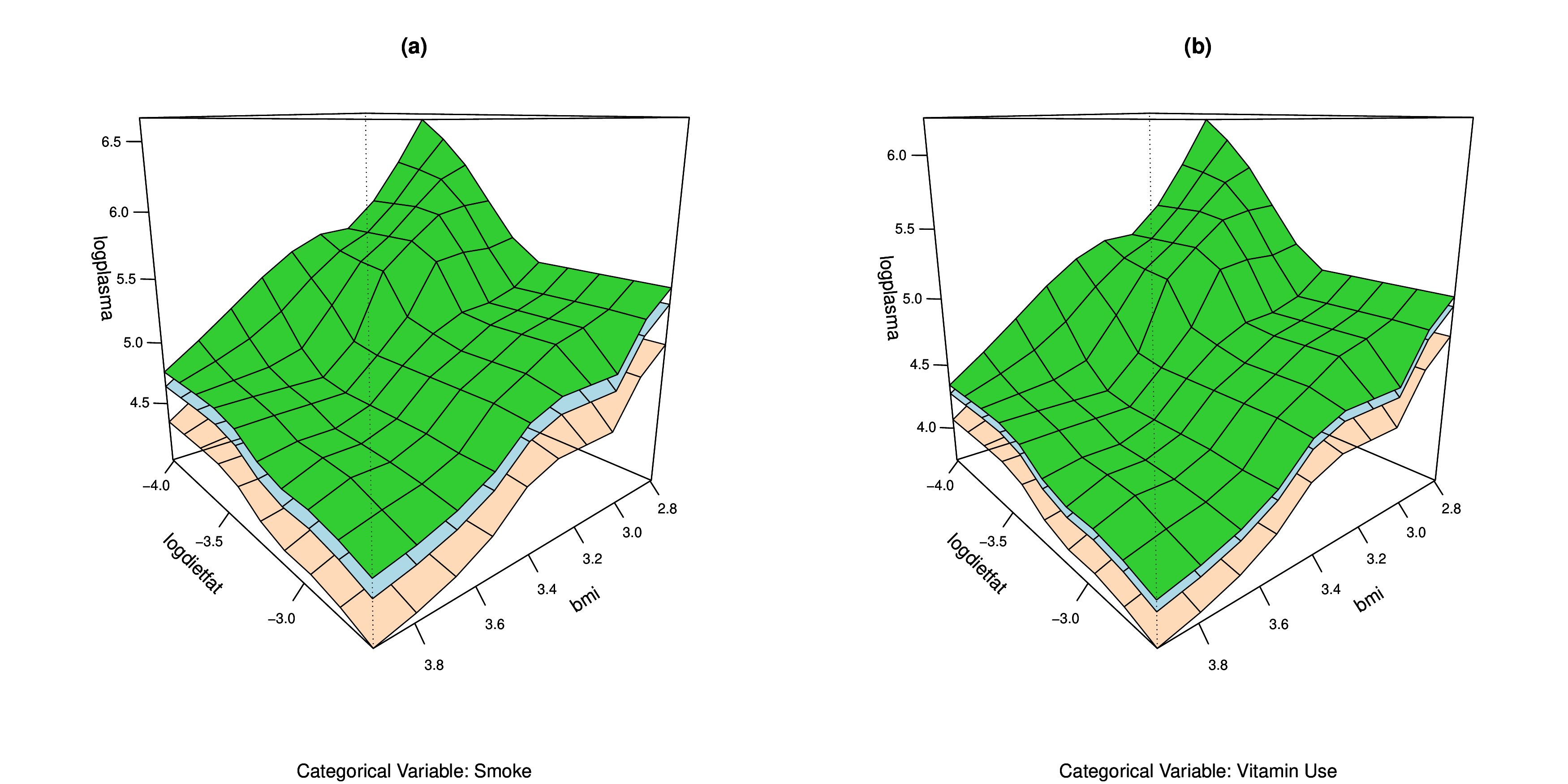}
\captionof{figure}{Demonstration of constrained regression using the {\tt wps} routine with the ``plasma'' data set. In each plot, the estimated surface is constrained to be doubly-decreasing in both predictors without the assumption of additivity. Parallel surfaces representing the effects of three levels of (a) ``smoke'' and  (b)  ``vituse,'' in an ascending order.}
\label{figure5}
\end{figure}
With $100$ simulations, the CIC value is $5.02$ for the doubly-decreasing model, and $4.97$ for the additive model in Section~\ref{sec52}; this is evidence that the additive model is adequate. 
\section[Discussion and comparison to similar packages]{Discussion and comparison to similar packages}  
\label{sec6}
The \proglang{R} package \pkg{scam} ({\bf s}hape {\bf c}onstrained {\bf a}dditive {\bf m}odel, \citet*{pya15}) uses penalized splines to fit shape-constrained model components.   The $P$-splines proposed in \citet*{eil96} are used with coefficients subject to linear constraints.  The shape options are similar to those in  \pkg{cgam}, but the back-fitting method for estimation in \pkg{scam} is slower than the single cone projection used in  \pkg{cgam}.   To compare the speeds for smooth fitting of regression functions, we simulated 1000 datasets from a regression model with three predictors, and fit isotonic additive models using \pkg{cgam} and \pkg{scam}.   Specifically, $x_{1i}$, $x_{2i}$, and $x_{3i}$, $i=1,\ldots,n$, were simulated independently and uniformly on $[0,1]$ along with independent standard normal errors $\ve_i$, $i=1,\ldots,n$,  then, $y_i = x_{1i} +x_{2i}^2 + x_{3i}^3+\ve_i$.  When $n=100$, the time used by {\tt cgam} and {\tt scam} are about $17$ and $611$ seconds, respectively.   When $n=500$, the time used by {\tt cgam} and {\tt scam} are about $58$ and $842$ seconds. The speed comparisons were made on a laptop with a $2.16$GHz dual-core Intel(R) Celeron(R) CPU.

\begin{figure}
\centering 
\includegraphics[height=3.0in,width=6.5in]{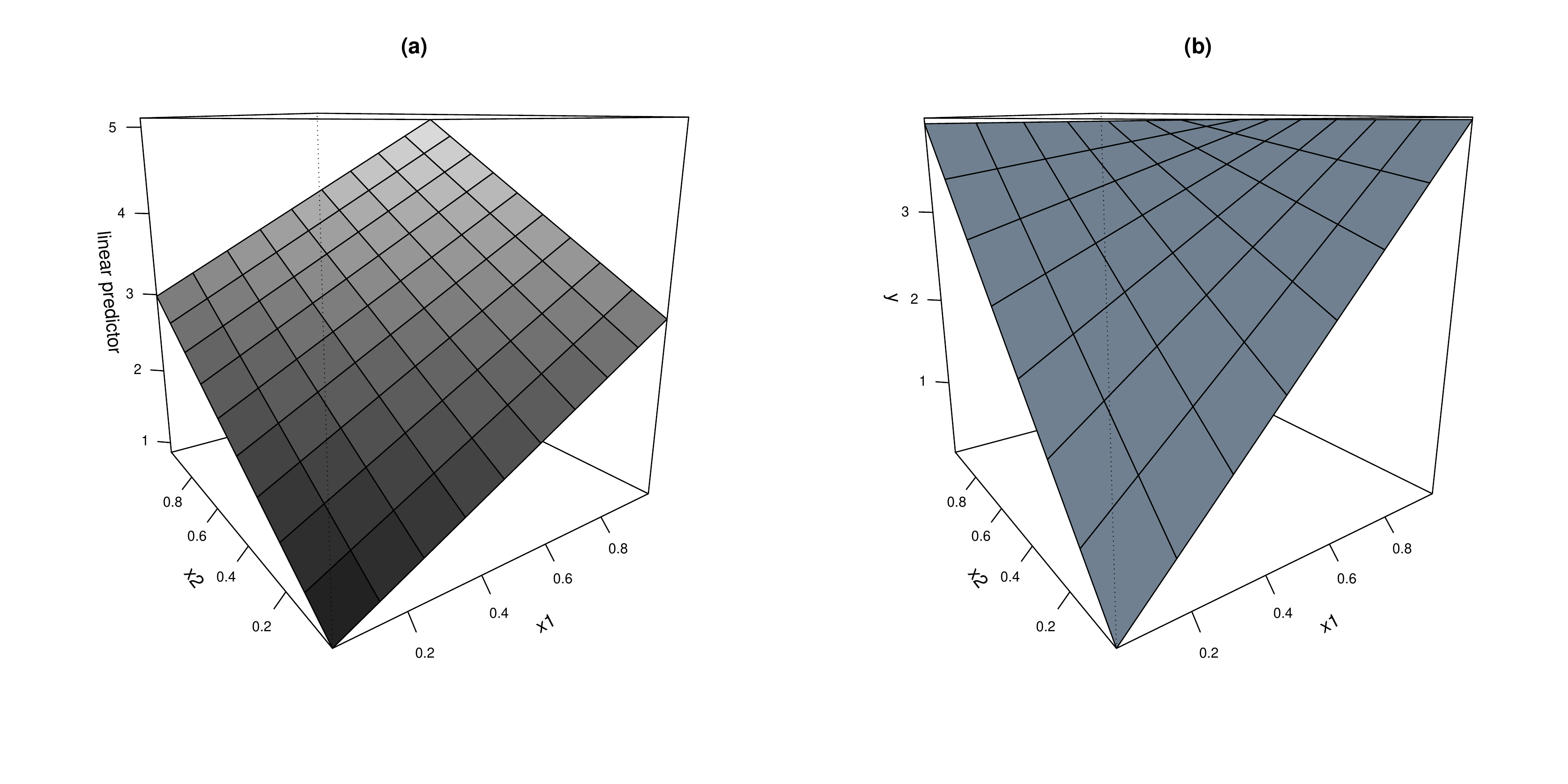}
\captionof{figure}{(a) estimated doubly-increasing regression surface using the {\tt scam} routine. (b) estimated doubly-increasing regression surface using the {\tt wps} routine.}
\label{fig:wps_scam}
\end{figure}

The \pkg{scam} package also has a routine to fit a bivariate isotonic regression function without additivity.  However, the constraints are sufficient but not necessary, and in fact severely over-constrain the fit.     In particular, it can not fit surfaces that are doubly monotone, but whose rate of increase is decreasing in at least one of the dimensions.  To demonstrate this,
we simulated from the regression surface $f(x)=4(x_1 + x_2 - x_1x_2)$ which is increasing over the unit square. The sample size $n=50$, each predictor is uniformly generated on the unit interval, and we choose not to add errors in this example. 
We use the default settings in \pkg{scam} and \pkg{cgam} to get doubly-increasing fits. An example of each method is shown in Figure \ref{fig:wps_scam} using the same data set.  

The package \pkg{scar} ({\bf s}hape {\bf c}onstrained {\bf a}dditive {\bf r}egression, \citet*{scarpaper}),  provides the maximum likelihood estimator of the generalized additive regression with shape constraints, but without smoothing. 

The packages \pkg{gam} and \pkg{mgcv} fit the generalized additive model, but without constraints. The main routine {\tt gam} in the \pkg{gam} package uses local regression or smoothing splines with the local scoring algorithm, which iteratively fits weighted additive models by back-fitting, while in the \pkg{mgcv} package, the main routine {\tt gam} only uses penalized smoothing splines and smoothness selection by criteria such as GCV is part of its model fitting. Besides, this {\tt gam} routine provides an option for modeling smooth interactions of any number of variables via scale invariant tensor product smooths.

The function specification of \pkg{cgam} is modeled after the popular \pkg{gam} function in the \pkg{mgcv} package with additional options for indicating shape. For example, a user may specify penalized smoothing in the \pkg{mgcv} package by
\begin{Schunk}
\begin{Sinput}
R> gam(y ~ s(x))
\end{Sinput}
\end{Schunk}
with no shape constraint, and choose spline-based regression with a shape constraint such as increasing by 
\begin{Schunk}
\begin{Sinput}
R> cgam(y ~ s.incr(x))
\end{Sinput}
\end{Schunk}
in the \pkg{cgam} package.

\end{document}